\documentclass{article}
\usepackage[a4paper]{geometry}                		
\usepackage{xcolor}
\usepackage{graphicx}
\usepackage[caption = false]{subfig} 
\usepackage[authoryear]{natbib}
\usepackage{bm}
\usepackage{amsmath}    
\usepackage{amssymb}
\usepackage{enumitem}
\usepackage{upgreek}
\usepackage{authblk}

\usepackage{tikz}
\DeclareRobustCommand\drawline[1]{%
  \raisebox{2pt}{\begin{tikzpicture}                                                           
    \draw[#1] (0pt,0pt) -- (15pt,0pt);                                       
  \end{tikzpicture}%
  }
} 
\DeclareRobustCommand{\mysquare}[1]{\tikz{\filldraw[draw=#1,fill=#1] (0,0) rectangle (0.2cm,0.2cm);}}

\def \E{{\rm E}}
\def \itH{{\bm H}}
\def \itH{{H}}
\def \bms{{\bm s}}
\def \bmx{{\bm x}}
\def \bmz{{\bm z}}
\def \iid{\stackrel{iid}{\sim}}
\newcommand{\mT}{\ensuremath{\mathsf{T}}}

\DeclareMathOperator*{\argmin}{arg\,min}
\DeclareMathOperator*{\argmax}{arg\,max}

\def \Var{{\rm Var}}
\newcommand{\MSE}{\text{MSE}}

\begin{document}

\title{Selecting a Scale for Spatial Confounding Adjustment} 

\author[1]{{Joshua P.} {Keller}}
\author[2]{{Adam A.} {Szpiro}}
\affil[1]{Colorado State University}
\affil[2]{University of Washington}

\renewcommand{\thefootnote}{}
\footnotetext{\noindent Corresponding author: Joshua P. Keller, Department of Statistics, Colorado State University, Fort Collins, CO USA.
Email: {joshua.keller@colostate.edu}.}

\date{}

\maketitle

\begin{abstract}
Unmeasured, spatially-structured factors can confound associations between spatial environmental  exposures and health outcomes. Adding flexible splines  to a regression model is a simple approach for spatial confounding adjustment, but the spline degrees of freedom do not provide an easily interpretable spatial scale. 
We describe a method for quantifying the extent of spatial confounding adjustment in terms of the Euclidean distance at which variation is removed.  We develop this approach for confounding adjustment with splines and using Fourier and wavelet filtering.
We demonstrate differences in the spatial scales these bases can represent and provide a comparison of methods for selecting the amount of confounding adjustment. We find the best performance for selecting the amount of adjustment using an information criterion evaluated on an outcome model without exposure.
We apply this method to spatial adjustment in an analysis of particulate matter and blood pressure in a cohort of United States women.

 \end{abstract}

\section{Introduction}
In environmental epidemiology,  we are often interested in making inference about the relationships between spatially-referenced exposures and spatially-referenced health outcomes. Spatial structure in exposures can arise from underlying environmental processes and human activities, while health outcomes can have spatial structure derived from both the exposure of interest and other spatially-varying factors. For example, coronary heart disease is associated with socioeconomic status \citep{DiezRoux2001}, which can have geographic structure and is associated with air pollution \citep{Brochu2011, Jerrett2004, Mensah2005}.  When not measured (or measured incompletely), these factors can cause \emph{unmeasured spatial confounding}, which is the inability to distinguish the effect of a spatially-varying exposure from residual spatial variation in a health outcome,  resulting in biased point estimates and standard errors \citep{Paciorek2010}.

Early discussion of spatial confounding in the literature includes \citet{Clayton1993}, who described the `confounding effect due to  location' in regression models for ecological studies, which they attributed  to unmeasured confounding factors. \citet{Clayton1993} advocated for the inclusion of a spatially-correlated error term in hierarchical models for spatial data \citep{Clayton1987, Besag1991} and claimed this would account for  confounding bias but might result in conservative inference.
Since then, the  approach  of adding spatially structured error terms has frequently been used in spatial  models for areal data \citep{Reich2006, Wakefield2007, Hodges2010, Hughes2013, Hanks2015, Page2017}.  In a causal inference framework, confounding due to geography has been recently addressed using spatial propensity scores \citep{Papadogeorgou2019, Davis2019}, however in those settings the exposure of interest was not explicitly spatial.

For point-referenced data, \citet{Paciorek2010} provided a rigorous discussion of spatial confounding and 
 the importance of the spatial scales of variability in the exposure and outcome.
\cite{Paciorek2010} demonstrated that reductions in bias could be obtained when the scale of unconfounded variability in the exposure, which he quantified by the spatial range parameter in a Matern covariance function,  is smaller than the scale of confounded variability. 

One approach in the literature for adjusting for spatial confounding at broad scales with point-referenced data is to estimate the coefficient of interest from a semi-parametric model that includes spatial splines \citep{Paciorek2010, Chan2015}.  This approach is a natural extension of time series studies, where flexible, one-dimensional basis functions can account for confounding due to unmeasured temporal variation \citep{Burnett1991,Schwartz1994, Dominici2003a, Dominici2004, SzpiroSheppard2014}.  Thin-plate regression splines (TPRS) \citep{Wood2003} are a commonly used basis and the amount of adjustment can be tuned using the degrees of freedom ($df$) parameter \citep{Paciorek2010}. However, particular values of $df$ do not have clear spatial scale or interpretation. Generalizing inference about associations between exposures and health outcomes is difficult when the extent of spatial confounding adjustment is not easily quantified.

\subsection{Air Pollution in the NIEHS Sister Study}
\label{sec:spatconf_motivatingexmp}
This work is motivated by an analysis of systolic blood pressure and fine particulate matter (PM$_{2.5}$) in the NIEHS Sister Study.  \citet{Chan2015} found that a 10 $\upmu$g/m$^3$ difference in long-term exposure to ambient fine particulate matter (PM$_{2.5}$) was associated with 1.4 mmHg (95\% Confidence Interval [CI]: 0.6, 2.3) higher systolic blood pressure (SBP) at baseline. 
To account for spatial confounding from unmeasured regional differences in socioeconomic and health patterns,  \citet{Chan2015} included TPRS with 10 $df$ in their model.

\begin{figure}[t]
\begin{center}
\includegraphics[width=0.9\textwidth]{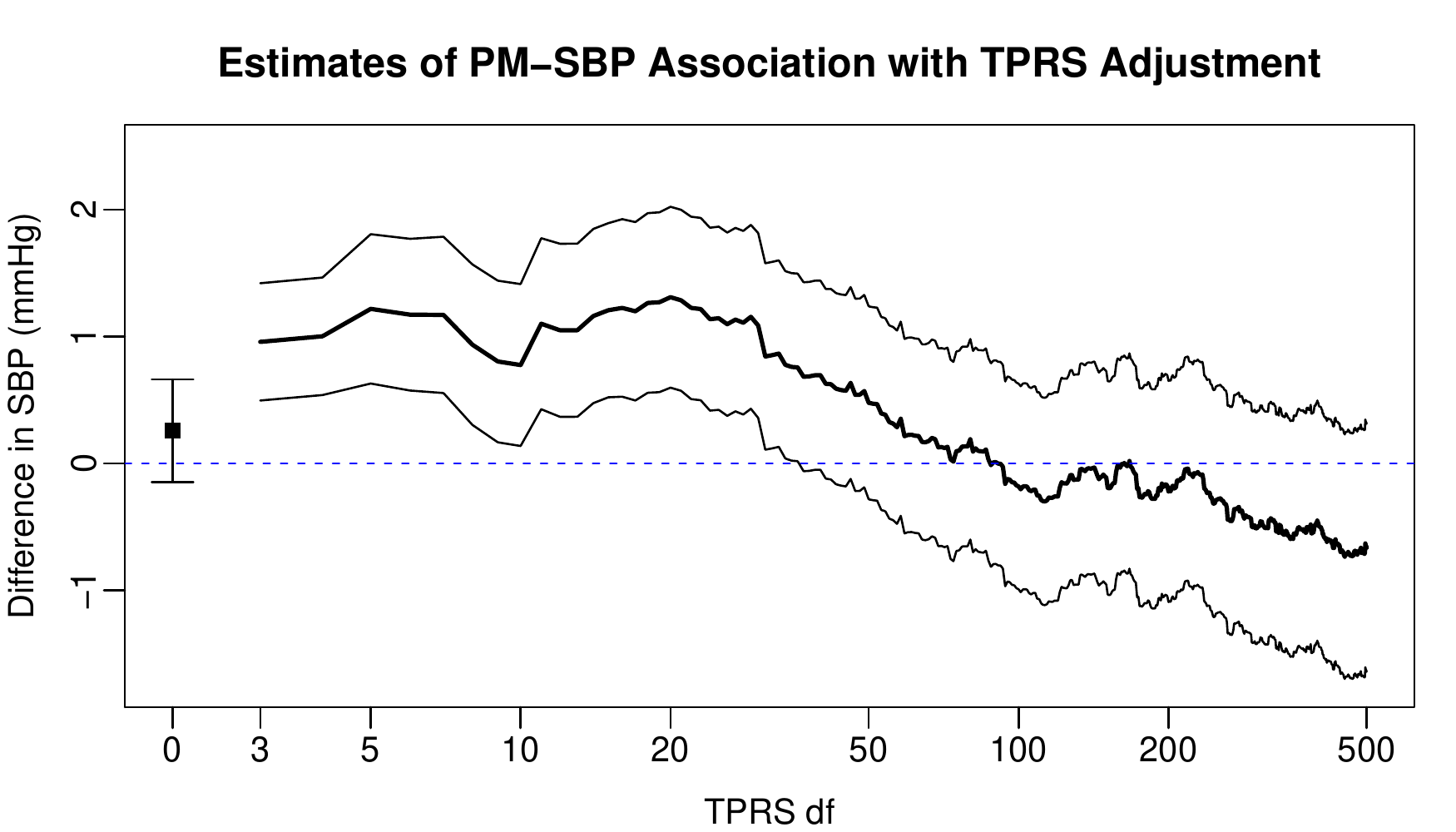}
\caption{Estimated difference in SBP (in mmHg) associated with a difference of 10 $\upmu$g/m$^3$ in annual average ambient PM$_{2.5}$,   when adjusting for TPRS with varying values of $df$ in the outcome model. The square marker (\mysquare{black}) at $df=0$ is the estimate without spatial adjustment and has error bars indicating a 95\% confidence interval.  The thick black curve (\drawline{ultra thick}) represents estimates for different choices of $df$ and the thin black curves (\drawline{}) represent point-wise 95\% confidence intervals.}
\label{fig:sister_tprsdf}
\end{center}
\end{figure}

Here we consider a re-analysis of this cohort using PM$_{2.5}$ exposures at  grid locations, rather than at subject residences, to accommodate the methods we describe below. 
We use predictions of the 2006 annual average ambient concentration from the universal kriging model of \citet{Sampson2013}, made on a  25km by 25km grid across the contiguous United States.
For each of the 47,206  Sister Study subjects, we assign exposure based upon the closest grid cell center.
  Using the same measured confounders as \citet{Chan2015},  but no spatial adjustment, we find that a difference of 10 $\upmu$g/m$^3$ in PM$_{2.5}$ is associated with 0.26 mmHg higher SBP (95\% CI: -0.14, 0.66). However, when TPRS with 10 $df$ are added to the model then the estimated difference in SBP is 0.77 mmHg (95\% CI: 0.14, 1.41). 
  Figure~\ref{fig:sister_tprsdf} shows the estimates for other amounts of adjustment. 
The change in the estimates as a function of $df$
 suggests that some form of spatial confounding is present, but it is not clear from Figure~\ref{fig:sister_tprsdf} on what scales the confounding is occurring,  how much adjustment should be done, and which estimate should be reported.

\subsection{Manuscript Outline}
In the remainder of this paper, we address the question of how to interpret and select the spatial scale  of adjustment.
In Section~\ref{sec:adjustment} we  introduce the statistical framework and  procedures for spatial confounding adjustment.
In Section~\ref{sec:spatial_scales} we describe three choices of spatial basis (TPRS,  a Fourier basis, and a wavelet basis) and present a method for assigning variation in these bases to a spatial distance. In Section~\ref{sec:choosing_m} we describe approaches for selecting the amount of adjustment and compare the approaches in simulations in Section~\ref{sec:simulations}. 
In Section~\ref{sec:sisterfullapp} we  apply the adjustment approaches to  the motivating Sister Study cohort. Section~\ref{sec:discussion} provides a concluding discussion.

\section{Methods for Spatial Confounding Adjustment}
\label{sec:adjustment}

\subsection{Statistical Framework}
\label{sec:framework}
We consider a cohort study of $n$ subjects with measured health outcomes $y_i$, exposures $x_i$, and observed confounder covariates $\bmz_i$. 
We partition the spatial domain into a rectangular grid  $\mathcal{S}$ with distinct locations $\{1, \dots, S\}$. 
Each subject is assigned to a location $\bms_i \in \{1, \dots, S\}$.  We assume 
that the subjects are distributed evenly across the locations; we discuss  non-uniform sampling of subject locations in Section~\ref{sec:pop_spatial_dist}.

Extending the framework of \cite{SzpiroSheppard2014} to a spatial setting, 
we assume that the exposure for each individual is measured without error and  takes the form
\begin{equation}
x_i = x(\bms_i) = g(\bms_i) + \xi(\bms_i),
\end{equation}
where $g(\bms_i)$ is a smooth spatial surface and $\xi_i$ is (fixed) residual spatial variation.  We decompose the observed  confounders  in a similar manner as $\bmz_i = \bm w(\bms_i) + \bm\zeta_i$, except that $\bm\zeta_i$ is not a function of space and is modelled as $N(0, \sigma_\zeta^2{\bm I})$. The stochasticity of $\bm\zeta$ arises from the sampling of subjects.

We assume that the observed health outcomes $y_i$ arise from the model
\begin{equation}
\label{eq:dgm}
y_i = \beta_0 + \beta x_i +  \bmz_i^\mT \bm\gamma + f(\bms_i) + \epsilon_i,
\end{equation}
where $f(\bms_i)$ is an unknown, fixed spatial surface and $\epsilon_i \iid N(0, \sigma_\epsilon^2)$.  The target of inference is the parameter $\beta \in \mathbb{R}$, which represents the association between exposure $x_i$ and the outcome $y_i$ \emph{conditional} on $\bm z$ and $f$. The $f(\bms_i)$ term represents unmeasured spatial variability in $y_i$, which we assume to be fixed and unknown.

In contrast to the approach taken by ``restricted spatial regression'', which targets the unconditional effect of exposure on outcome \citep{Hodges2010,Hughes2013}, our parameter of interest here is the exposure-outcome association conditional on the unmeasured confounder $f$. In observational air pollution epidemiology studies, the conditional association is targeted in order to provide information that is generalizable beyond the specific time and location of the study cohort \citep[e.g.][]{Peng2006}.

To model the spatially-varying terms in \eqref{eq:dgm}, we employ a set of hierarchical spatial basis functions $\{h_1(\cdot), h_2(\cdot), \dots. \}$ that are in order of non-increasing resolution.
 For compactness of notation, let $\itH_m  = \begin{bmatrix} \bm h_1 & \dots & \bm h_m\end{bmatrix}$ denote the $n \times m$ matrix of the first $m$ basis functions  evaluated at the $n$ subject locations.
Following \citet{Dominici2004} and \citet{SzpiroSheppard2014}, we decompose $g(\cdot)$, $\bm w(\cdot)$, and $f(\cdot)$ as:
  \begin{align}
\notag   g(\bms_i)  &= \sum_{k=1}^{m_1} \alpha_kh_k(\bms_i)  = \bm e_i^\mT\itH_{m_1}\bm\alpha\\
  \label{eq:gwf_basis} w_j(\bms_i) & = \sum_{k=1}^{m_2} \delta_{jk}h_k(\bms_i)  = \bm e_i^\mT\itH_{m_2}\bm\delta_j\\
\notag   f(\bms_i) &=  \sum_{k=1}^{m_3} \theta_kh_k(\bms_i) = \bm e_i^\mT\itH_{m_3}\bm\theta,
  \end{align}
  where $\bm e_i$ is the unit vector with 1 as the $i$th element.
We assume 
$m_1 > m_3$. This assumption means that there is finer scale variability in the exposure surface than in the unobserved confounding surfaces, which \citet{Paciorek2010} identified as a necessary condition to achieve reductions in bias via adjustment.
In Section~\ref{sec:spatial_scales}, we consider three different choices of $h_j$: a regression spline basis, a Fourier basis, and a wavelet basis.

The ordinary least squares estimator from the model $\E[y_i] = {\beta}_0 + {\beta}_1x_i + {\gamma} \bmz_i$
 will be biased due to the correlation of the observed $g(\bms_i)$ and  $\bm w(\bms_i)$ with the unobserved $f(\bms_i)$. In the literature, this correlation is sometimes referred to  as
 \textit{concurvity} \citep{Hastie1990, Ramsay2003,Ramsay2003a}.
 We describe two strategies for adjusting for this correlation in order to eliminate the confounding bias.
  
  \subsection{Adjustment in Outcome Model}
\label{sec:adjust_health}
A straightforward way to adjust for the unmeasured spatial structure $f(\bms_i)$ is to fit the semiparametric regression model
\begin{equation}
\label{eq:adj_in_health_model}
\E[y_i] = \breve{\beta}_0 + \breve{\beta}_1x_i +  \bmz_i^\mT\breve{\gamma} + \itH_m\breve{\bm \theta},
\end{equation}
for a chosen value of $m$.
This is a direct extension of confounding-adjustment approaches used in time series methods \citep{Dominici2004,Peng2006} and was the approach taken by \cite{Chan2015} and in Figure~\ref{fig:sister_tprsdf}. For basis functions such as TPRS that are easily computed, fitting model \eqref{eq:adj_in_health_model} in standard statistical software is straightforward. 

If $m \ge m_3$, then the ordinary least squares estimator $\widehat{\breve{\beta}}_1$ from model \eqref{eq:adj_in_health_model} will be unbiased, because the inclusion of $\itH_m$ fully adjusts for variation in its column space, which includes $f$. 
If $m < m_3$, then $\widehat{\breve{\beta}}_1$ will remain biased because of correlation between $x(\bms)$ and the $m_3 - m$ terms of $f(\bms)$ (from the decomposition in \eqref{eq:gwf_basis}) that are not within the column space of $\itH_m$.

\subsection{Pre-adjusting Exposure}
\label{sec:preadjust}

A second approach to confounding adjustment is to first `pre-adjust' the  exposure, to remove variation in the exposure that is correlated with the confounding surface.  In comparison to the semi-parametric adjustment approach described above, the pre-adjustment approach does not necessarily require the explicit construction of $\itH_m$, which allows for additional choices of spatial basis.
\cite{SzpiroSheppard2014} outlined how preadjustment can be done for cohort studies with time series exposures, and here we present an extension of that approach to spatial settings. This approach can be considered a form of spatial filtering \citep[e.g.][]{Haining1991, Tiefelsdorf2007}

We decompose the vector of observed exposures $\bmx = \begin{bmatrix} x(\bms_1) & \cdots &x(\bms_n)\end{bmatrix}^\mT $ into two components, $\bmx_1$ and $\bmx_2$, such that $\bmx_2$ is orthogonal to $\itH_m$ for some $m$ (we discuss methods for selecting $m$ in Section~\ref{sec:choosing_m}). 
In settings when $\itH_m$ can be explicitly computed, this decomposition can be accomplished by taking $\bmx_1$ to be the projection of $\bmx$ onto $\itH_m$ ($\bm x_1 = \itH_m(\itH_m^\mT\itH_m)^{-1}\itH_m^\mT\bmx$) and $\bmx_2$ to be its complement ($\bm x_2 = \bmx - \bmx_1)$.
When it is not practical to explicitly construct $\itH_m$, as in the Fourier and wavelet approaches we discuss in Section~\ref{sec:spatial_scales}, we use a filtering procedure in the frequency domain to achieve this decomposition.

Once $\bmx$ has been partitioned into $\bmx_1$ and $\bmx_2$,  we fit the model
\begin{equation}
\label{eq:preadj_model}
y_i = \tilde{\beta}_0 + \tilde{\beta}_1x_{1i} + \tilde{\beta}_2x_{2i}+ \tilde{\gamma} \bmz_i + \tilde{\epsilon}_i
\end{equation}
where $\tilde\epsilon_i = \epsilon_i + f(s_i)$ and $\tilde\beta_1, \tilde\beta_2 \in \mathbb{R}$.  Because $\sum x_1x_2 = 0$, the bias of $\hat{\tilde{\beta}}_2$ is proportional to
\begin{equation}
\label{eq:beta2bias}
\sum x_2f\left(\sum x_1z^2 - \sum x_1^2\sum z^2\right) + \sum x_2z\left(\sum x_1^2\sum zf - \sum x_1z\sum x_1f\right),
\end{equation}
where  the summations are over $i$. If $m \ge m_3$, then $\bmx_2$ is by construction orthogonal to $\bm f$ and so $\sum x_2 f = 0$. If $m \ge m_2$, then $\bmx_2$ is orthogonal to $\bm z$ and so $\sum x_2z=0$. Under these two conditions, the bias of $\hat{\tilde{\beta}}_2$ given in \eqref{eq:beta2bias} is zero.
If either $m < m_2$ or $m < m_3$, then $\hat{\tilde{\beta}}_2$ will be biased.  Forcing $f(s_i)$ into the error term creates correlation between the $\tilde\epsilon_i$, but this can be accounted for using `sandwich' standard error estimates \citep{White1980}.

\subsection{Non-Uniform Spatial Distributions of People}
\label{sec:pop_spatial_dist}

So far we have assumed  that all locations $s \in \mathcal{S}$ have an observed subject and that the number of subjects was that same at each location. This assumption was necessary to achieve the orthogonality of the pre-adjusted exposure ($\bmx_2$) and the basis functions matrix $H_m$ (and thus $\bmx_1$, $\bmz$, and $\bm f$). 
Adjustment in the outcome model or exposure pre-adjustment in which $H_m$ can be explicitly constructed do not require that the locations be spatially uniform. However, pre-adjustment using the filtering approaches does require that all locations be observed the same number of times.

To implement pre-adjustment in the presence of duplicate locations, first the duplicated locations are removed (and unobserved locations added) to create the vector $\bm x'$. The filtering pre-adjustment is done on $\bm x'$ to generate $\bm x_1'$ and $\bm x_2'$. By construction $\bm x_2'^\mT \bm h_j' = 0 $, where $\bm h_j'$ is the $j$th basis function evaluated at all locations in $\mathcal{S}$ once.
The duplicated locations are then added back in by repeating the relevant entries in $\bmx_2'$ to form $\bm x_2$ (and similarly for $\bm x_1$). However, the duplicated locations means that in general $\bm x_2^\mT \bm h_j \ne 0$. To recover the orthogonality necessary to eliminate the bias term \eqref{eq:beta2bias}, the outcome model can use re-weighting.

Suppose that subject locations are distributed according to a spatial distribution $\Phi(\bms)$, which has positive support across all of $\mathcal{S}$. 
To recover orthogonality of  $\bmx_2$ and $\bm h_j$ ($j=1, \dots, m$), we reweight the study population by its spatial distribution. Specifically, we fit a weighted least squares (WLS) version of \eqref{eq:preadj_model} where the observation weights are proportional to the inverse of $\frac{d}{d \bms} \Phi(\bms) = \phi(\bms)$.  
This assures orthogonality of $\bmx_2$ and $\bm h_j$ as $n\to \infty$, since
\begin{equation*}
 \frac{S}{n} \sum_{i=1}^n \frac{1}{\phi(\bms_i)}x_2(\bms_i)h_j(\bms_i) \rightarrow \sum_{\bms \in \mathcal{S}} \phi(\bms) \left(\frac{1}{\phi(\bms)} x_2'(\bms)h_j'(\bms)\right)  = \sum_{\bms \in \mathcal{S}} x_2'(\bms) h_j'(\bms) = \bm x_2'^\mT\bm h_j' = 0.
\label{eq:reweighting}
 \end{equation*}

This WLS approach essentially downweights the contribution of people living in highly-populated areas and gives added weight to those who live in more sparsely populated areas.  The parameter being estimated remains the same. For applications in air pollution epidemiology, the correlation between high exposures and population density means that this has the effect of weighting subjects with different exposure levels differently. This is similar in spirit to propensity score weighting, in which groups are re-weighted by their probability of exposure in order to remove bias due to confounding \citep{Stuart2010}.

\section{Interpreting Spatial Scales of Adjustment}  
\label{sec:spatial_scales}

Here we present specific details for three different choices of spatial basis functions: thin-plate regression splines, Fourier basis, and wavelet basis.  We describe how to implement adjustment for each basis and how to interpret the amount of adjustment in terms of physical distance. 
We will describe methods for selecting the amount of adjustment in Section~\ref{sec:choosing_m}.

\subsection{Effective Bandwidth $\hat k$}
For each basis, we describe how to relate the amount of confounding adjustment (i.e.  $m$) to a physical spatial scale. We call this scale, denoted by $\hat k$, the \emph{effective bandwidth} of the adjustment. Heuristically, the effective bandwidth is the width of the smoothing kernel induced by the choice of basis function.   This allows for adjustment using $m$ basis functions to be interpreted as adjusting for confounding by smoothing at $\hat k$ distance. In the context of particular application,  $\hat k$ may represent a national, regional, or metropolitan scales of variation.  

\subsection{Thin-plate regression splines}
\label{sec:tprs}
Thin-plate regression splines (TPRS) are low-rank approximations to  thin-plate smoothing splines  \citep{Wood2003}.  The full-rank thin-plate splines are derived from a set of radial basis functions, and TPRS achieve computational benefits by using a truncated eigenbasis  to approximate the radial basis. TPRS are based upon the spatial locations of observed points, eliminating the need for knot-selection.

For problems of inference, unpenalized TPRS are indexed by the degrees of freedom ($df$), which controls the dimension of the basis and  can be fixed to yield a set of hierarchical basis functions $h_1(s), h_2(s),  \dots, h_{df}(s)$ where higher values of $df$ correspond to adjustment at a finer scale.   The \texttt{mgcv} package in R \citep{Wood2003} makes explicit construction of the TPRS basis straightforward, so both the semi-parametric adjustment and pre-adjustment approaches can be used with this basis.

We propose the following procedure for computing the effective bandwidth for a chosen number of TPRS basis functions. The smoothing induced by TPRS can be represented in the smoothing matrix  $ S = \itH_{df}(\itH_{df}^\mT\itH_{df})^{-1}\itH_{df}^\mT$.  Unlike the frequency-based methods discussed below, the amount of smoothing, which is given by the columns of $ S$, varies at each location due to the location of neighboring points, which is also impacted by the geometry of the domain. We compute $\hat k$ by finding the  distance at which the ``average'' value of the smoothing matrix values first cross zero. For each observation $i$, we first compute the vector of distances $\bm d_i$ to all other points. We then fit a loess smooth to $\bms_i$, the $i$th column of $S$, as a function of $\bm d_i$.  The procedure is repeated across all locations, and the pointwise median of the loess smooths computed. The distance at which this median smooth first crosses zero is $\hat k$. A detailed description of this procedure is provided in Supplementary Material Section A. The publicly-available R package  \texttt{spconf} implements this procedure for any provided spline or smoothing matrix values.

\subsection{Fourier decomposition and Frequency-based filtering}
\label{sec:fourier}
A second choice for  basis $\{h_j\}$ is the set of sinusoidal Fourier basis functions.  Without loss of generality, let $u=0, \dots, M-1$ and $v=0, \dots, N-1$  denote coordinates of $\mathcal{S}$. 
The value of the function $g(u, v)$ can be written as a linear combination of sine and cosine functions, whose coefficients are determined by the Discrete Fourier Transform (DFT) of $g(u, v)$ \citep{Gonzalez2008}.  
Spectral methods have been widely used in spatial statistics, often as a means to approximate covariance functions \citep[e.g.][]{Guinness2017, Royle2005,Fuentes2007}.
  
Allowing for ties as necessary, we can use the spectral coordinates $(p, q)$ to order the spatial basis functions by \emph{effective frequency}: $\omega_{(p, q)} = \sqrt{\left(\frac{p}{M}\right)^2 + \left(\frac{q}{N}\right)^2}$  \citep{Burger2009}.
Because each basis function is oriented to a particular angle, there may be multiple basis functions with the same frequency $\omega$ but different orientation. 

To avoid explicit construction of all $MN$ basis functions, we apply the pre-adjustment approach described in Section~\ref{sec:preadjust} for a selected frequency $\omega$. We decompose  $\bmx$ into its projection onto $H_{\omega}$ and corresponding complement by applying a high-pass filter in the frequency domain.  The values of the filter 
 are
\begin{equation}
F_{\omega}(p, q) = \begin{cases} 1 & \text{ if } \omega_{(p, q)}> \omega \\ 0 & \text{ if } \omega_{(p, q)} \le \omega \end{cases}.
\end{equation}
To implement pre-adjustment, we first compute the DFT $\mathcal{X}(p,q)$ of $\bm x'$, which is the gridded exposure surface $\bmx(\bms)$ with duplicated locations removed.  We then define $\mathcal{X}_{\omega}$ to be the element-wise product of $\mathcal{X}$ and $F_{\omega}$. The inverse DFT of $\mathcal{X}_{\omega}$ provides $\bmx_2'$. Duplicated locations are assigned the same value from $\bmx_2'$, yielding $\bmx_2$.

One interpretation of removing variation described by frequencies $\omega$ and lower is that it removes periodic variation with periods $1/\omega$ and higher.  %
The analogue of $\hat k$ for the Fourier basis is the effective bandwidth of the kernel smoother in the spatial domain that corresponds to the frequency filter $F_{\omega}$. 
This smoother is given by the inverse Fourier transform of $F_{\omega}$, which is analytically a 'sinc' function. However, edge effects from finite grids mean the  inverse DFT of $F_{\omega}$ does not exactly follow the expected analytic form. Therefore, we compute the effective bandwidth $\hat k$ of $F_{\omega}$ empirically in a manner similar to the approach for TPRS: Let $F'_\omega(u, v)$ denote the inverse DFT of $F_{\omega}$. We take $\hat k$ to be the minimum value of $\sqrt{u^2 + v^2}$, scaled by grid size, for which $F'_\omega(u, v)=0$. 

\subsection{Wavelet Basis and Thresholding}
\label{sec:wavelets}
A third choice of spatial basis  $\{h_j\}$ is a set of  wavelets. Like Fourier basis functions, wavelets are localized in frequency, but they are also localized in space \citep{Nason2008}. This means that wavelets can compactly describe variation at different frequencies in different areas of a spatial domain. There are many different sets of wavelets, and here we use the smooth Daubechies wavelets \citep{Daubechies1988}.
 Wavelet basis functions are indexed by \emph{level}, based upon successive halving of the spatial domain. Within each level $L$ (and thus, within each frequency $2^{L}$), there are multiple wavelet basis functions with different orientations and spatial positions. The effective bandwidth that corresponds to each level is $\hat k = 2^{-L}$.
 
 Multi-resolution approaches such as wavelets have recently been used to quantify spatial variation of air pollution exposures \citep{Antonelli2017}. However, \cite{Antonelli2017} only described exposures with variation at ``low" and ``high'' frequencies, and did not attempt to use wavelets for explicit confounding adjustment nor did they provide a specific distance to these labels.  De-correlation of the exposure and outcome via wavelets was presented in an ecological context by  \citet{Carl2008}. However,  they applied the thresholding to the exposure and the outcome jointly and performed regression on the wavelet coefficients, which is not practical for the large grids and when we have other confounders in the model.
 
 For finite data on a discrete grid, the Discrete Wavelet Transform (DWT) maps any surface to a set of wavelet coefficients.  
 Unlike the DFT, the DWT requires that the grid have length equal to a power of two.  Data on a non-square grid can be embedded within  a larger grid with dyadic dimension to apply the DWT. We use an implementation of the DWT available in the R package \texttt{wavethresh} \citep{Nason2008}.

To pre-adjust exposure using wavelets, we apply a filtering approach similar to that used for a Fourier basis. We first compute the DWT $\mathcal{W}$ of $\bmx'$. We then threshold  all coefficients at levels $\ell = 0, \dots, L$ to get a modified wavelet transform $\mathcal{W}_L$.
 We then apply the inverse wavelet transform to $\mathcal{W}_L$ to get the pre-adjusted exposure $\bmx_2'$.
Although we use a global threshold of all wavelet coefficients up to a particular level, the extent of thresholding could be varied across the exposure domain if desired.


\section{Selecting the amount of adjustment}
\label{sec:choosing_m}

A critical question for data analysis is how much adjustment to do or, equivalently, what value of $\hat k$ or $m$ should be chosen. In both approaches presented in Section~\ref{sec:adjustment}, the estimates of $\beta$ are unbiased only if $m \ge m_3$ (with an additional condition of $m \ge m_2$ for the pre-adjustment approach) and if the basis for adjustment is correctly chosen. 
But in most practical settings, the true values of $m_1$, $m_2$, and $m_3$ and the correct choice of basis $\{h(\cdot)\}$ are unknown. 
In this section, we describe different approaches to choosing the amount of adjustment ($m$ or $\hat k$).

If there is specific  knowledge of assumed unmeasured confounders or other external, content-area knowledge, then $m$ could be selected \emph{a priori}. Exact knowledge of $m_3$ is unlikely to be known, but a choice of $m$ or $\hat k$ might be based upon a combination of known scale of variation in the exposure (e.g. if it is predicted from a spatial model with known parameter values) and the approximate scale of the possible confounders (e.g. regional variation in socioeconomic status). This choice might also be influenced by estimates of $m_1$ and $m_2$, which could be estimated from the data. While an \emph{a priori} choice of $m$ may not necessarily lead to an unbiased estimate, it does provide a way to pre-specify the extent and interpretation of adjustment, without risking overfitting the model to the outcome data $\bm y$.

In the absence of external knowledge, the amount of adjustment $m$ can be selected in a data-driven manner. Estimating $m_3$ directly is challenging, since it requires knowing the amount of variation in $y$ that is due to $\bm x$ and $\bm z$--exactly what we are trying to estimate. However, measures of model fit (for either \eqref{eq:adj_in_health_model} or \eqref{eq:preadj_model}) can be used to calculate the scale of spatial variation in $\bm y$. Specifically, after fitting the outcome model with a range of choices for $m$, the model that minimizes the Akaike Information Criterion (AIC) \citep{Akaike1973} or Bayesian Information Criterion (BIC)  \citep{Schwarz1978} can provide a selection of the model that best fits the data. However, selecting the model that best fits $\bm y$ is not guaranteed to reduce error in the estimate of $\beta$.

A variation of this procedure for estimating the scale of relevant spatial variation in $\bm y$ is to fit outcome models that include the adjustment basis but do not include the exposure.  The amount of adjustment that leads to the smallest AIC or BIC can be chosen as $m$ for the primary model. Observed confounders can either be included in the model, or first projected out from $\bm y$. Here we consider the former approach and refer to is as the  ``AIC-NE'' approach for choosing $m$, respectively, where ``NE'' stands for ``No Exposure''.  The BIC alternative is named in the analogous manner.
The rationale of this approach is that if the unmeasured confounders have a strong impact on the values of $\bm y$ and the association of interest ($\beta$) is relatively weak (a relatively common occurrence in environmental epidemiology), this approach can yield a choice of $m$ without overfitting the relationship between $\bm y$ and $\bm x$. Because it involves fitting a model with the outcome $\bm y$, this approach is mostly limited to adjustment bases such as TPRS that can be explicitly computed. If all locations on the grid had a single outcome value, then this approach could be used for Fourier and wavelet filtering, but that setting is unlikely to occur in practice. 

An approach  that targets estimation error directly is to choose $m$ to minimize the estimated MSE of $\hat{\beta}(m)$.  Let $m'$ be some large value for which it is assumed the estimator $\hat{\beta}$ is unbiased. Then estimate the bias of $\hat{\beta}(m)$ as $\hat{\beta}(m) - \hat{\beta}(m')$ and estimate $\Var(\hat{\beta}(m))$ via the `sandwich' estimator \citep{White1980} for the fitted model.  Together these provide a method for selecting $m$:
\begin{equation}
\label{eq:pick_m_mse}
\hat{m}_{MSE} = \argmin_{\tilde{m}} \widehat{\MSE}(\hat{\beta}(\tilde{m})) = 
\argmin_{\tilde{m}} \left\{\left[\hat{\beta}(\tilde{m}) - \hat{\beta}(m')\right]^2 + \widehat\Var(\hat{\beta}(\tilde{m}))\right\}
\end{equation}
This approach, however, has the clear drawback that $\hat{\beta}(m')$ may not be an unbiased estimator (see Section~\ref{sec:overadj}).

Approaches that are more \emph{ad hoc} are also possible. One such option is to choose $m$ by selecting a point right after a ``knee'' in the plot of $\hat{\beta}$ against $m$ (for example, Figure~\ref{fig:sister_tprsdf}). While visually intuitive, this approach requires a cumbersome formal definition. We define $\hat{m}_{knee}$ to be:
\begin{equation}
\label{eq:pick_m_knee}
\hat{m}_{knee} = \argmin_{\tilde m} \left\{|D^1(\hat{\beta}(\tilde m))| \Big| |D^1(\hat{\beta}(\tilde m))| < |D^1(\hat{\beta}(\tilde m + 1)), \tilde{m}  \in \mathcal{M}\right\},
\end{equation} 
where  $D^k(\cdot)$ is a $k$-order difference operator and $\mathcal{M} = \{m |  m >  \argmax_\omega D^2(\hat{\beta}(\omega))\}$.
The set $\mathcal{M}$  limits to values of $m$ that are beyond the largest second-order difference, an approximation of curvature, in the $\hat{\beta}(m)$ sequence. Within this set, the value of $m$ that gives the first minimum in the first-order differences is selected.  Importantly, this approach does not account for differences in the resolution of $m$ and can be greatly impacted by the noise in the finite differences.

\subsection{Impact of over-adjustment}
\label{sec:overadj}
If the choice of basis functions exactly matches the underlying confounding surface $f(s)$, then sufficiently rich adjustment will remove bias. However, if the choice of basis functions does not match the underlying confounding surface $f(s)$, then residual bias may remain regardless of the choice of $k$. This could lead to increases in bias, since the adjustment basis ($H_m$) then becomes a near-instrumental variable (IV) that is highly correlated with the exposure $x(s)$ but only weakly related to the outcome $y$. Adjustment for a near-IV in contexts with residual confounding can amplify bias from the residual confounders at a rate greater than any bias reduction due to confounding by the near-IV \citep{Pearl2011}. This suggests that the MSE approach presented above may perform quite poorly if the choice of basis functions is incorrect.


\section{Simulations}
\label{sec:simulations}
\subsection{Primary Setup}
We conducted a set of simulations to compare the different adjustment approaches, demonstrating both different choices of spatial basis and the different methods for selecting the scale of adjustment.
The data for the simulations are created at points $(u,v)$ lying on a $512 \times 512$ grid over the unit square $[0, 1) \times [0, 1)$.  
For each simulation, we constructed a fixed exposure surface $x(u, v)$ and a fixed unmeasured confounder surface $f(u, v)$ on this grid. 

We considered six different unmeasured confounder surfaces, $f_1, \dots, f_6$, constructed to have ``large'' and ``fine'' scale variation for each of three choices of basis: TPRS, sinusoidal functions, and a spatial Gaussian process (GP). Table~\ref{tab:simf} summarizes these surfaces, and mathematical detail on their definitions is provided in the Supplemental Material, Section B.1. 

\begin{table}
\caption{Description of the different confounder surfaces in the simulations.
\label{tab:simf}}
\centering
\fbox{
\begin{tabular}{ll}
Surface & Description\\
\hline
$f_1$ & TPRS with 10 df\\ %
$f_2$ & TPRS with 50 df\\ %
$f_3$ & Sinusoidal up to frequency $\sqrt{40}$\\ %
$f_4$ & Sinusoidal up to frequency $\sqrt{500}$\\ %
$f_5$ & Exponential GP with range 0.5 \\ %
$f_6$ & Exponential GP with range 0.15\\ %
\hline
\end{tabular}
}
\end{table}

We constructed the exposure surface as $x(u, v) = \theta f(u, v) + g(u, v)$,  where $g(u, v)$ is a fixed realization of a Gaussian process with exponential covariance structure.
This structure for $x$ allows for the amount of correlation between $x$ and $f$ to be controlled, without requiring that they are generated in the same manner (e.g. correlated Gaussian process as used by \cite{Page2017}).
 For Simulations 1 and 2, we considered two values for the range parameter in the covariance function used to generate $g(u,v)$: 0.05 and 0.50, respectively. The value of 0.05 results in smaller scale spatial variation than the candidate confounders, which is needed to eliminate bias.  The value of 0.50 results in spatial variation at a coarser scale than some of the confounder surfaces, which may lead to the persistence of bias but is also a situation that could reasonably arise in practice.  The parameter $\theta$ controls the correlation between $x$ and $f$ and thus the amount of bias due to unmeasured confounding. Because the correlations between $g(u, v)$ and $f(u, v)$ differ for each choice of $f(u, v)$, we calculated $\theta$ so that the amount of bias was equal for all confounder surfaces, which makes comparison of the results more straightforward. Plots of the surfaces $g, f_1, \dots, f_6$ and the formula for $\theta$ are provided in Supplemental Material Section B.2. The values of $x(u, v)$ and $f(u, v)$ were standardized to have mean zero and unit variance.

For each simulation replication, the outcome was constructed as
$y(u,v) = \beta x(u, v) +  f(u, v) + \epsilon(u, v),$
 where $\epsilon(u, v) \iid N(0, 16)$.  This setup is designed to reflect a feature of the Sister Study application in that the residual variation in the outcome, even after accounting for unmeasured confounding, is large relative to the exposure variation.
 We redrew a sample of  $n=2,000$ observation locations (selected from the grid of points) for 1,000 replications of each simulation.
For Simulations 1 and 2, we set $\beta = 1$ and chose $\theta$ such that the uncorrected bias would be 0.2. For each simulation we also conducted a variation with no effect ($\beta = 0$) to estimate Type 1 error at the nominal $\alpha=0.05$ level.
 
We compare estimates from four sets of models: 
 (i) a model with semi-parametric adjustment via TPRS  in the outcome model and models with pre-adjustment of exposure (ii) by TPRS (at observed locations),  (iii) via a high-pass Fourier filter, and (iv) via wavelet thresholding. We present estimates for a sequence of adjustment amounts and using the selection methods from Section~\ref{sec:choosing_m}.

\subsection{Effective bandwidth } 
\begin{figure}
\begin{center}
\subfloat[\label{fig:eff_bandwidth_tprs_unit_square}]{
\includegraphics[width=0.45\textwidth]{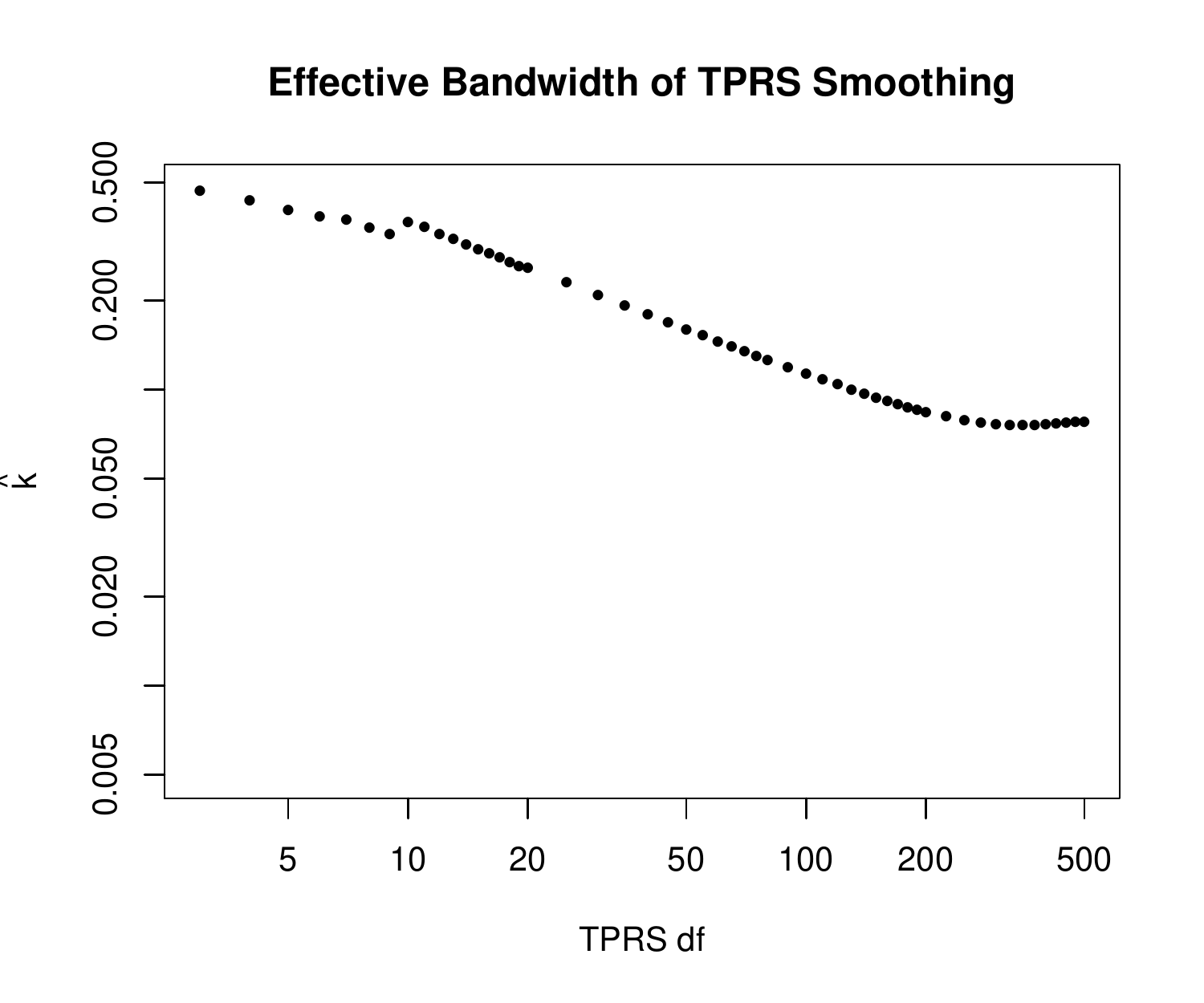}
}
\subfloat[\label{fig:eff_bandwidth_hpf_unit_square}]{
\includegraphics[width=0.45\textwidth]{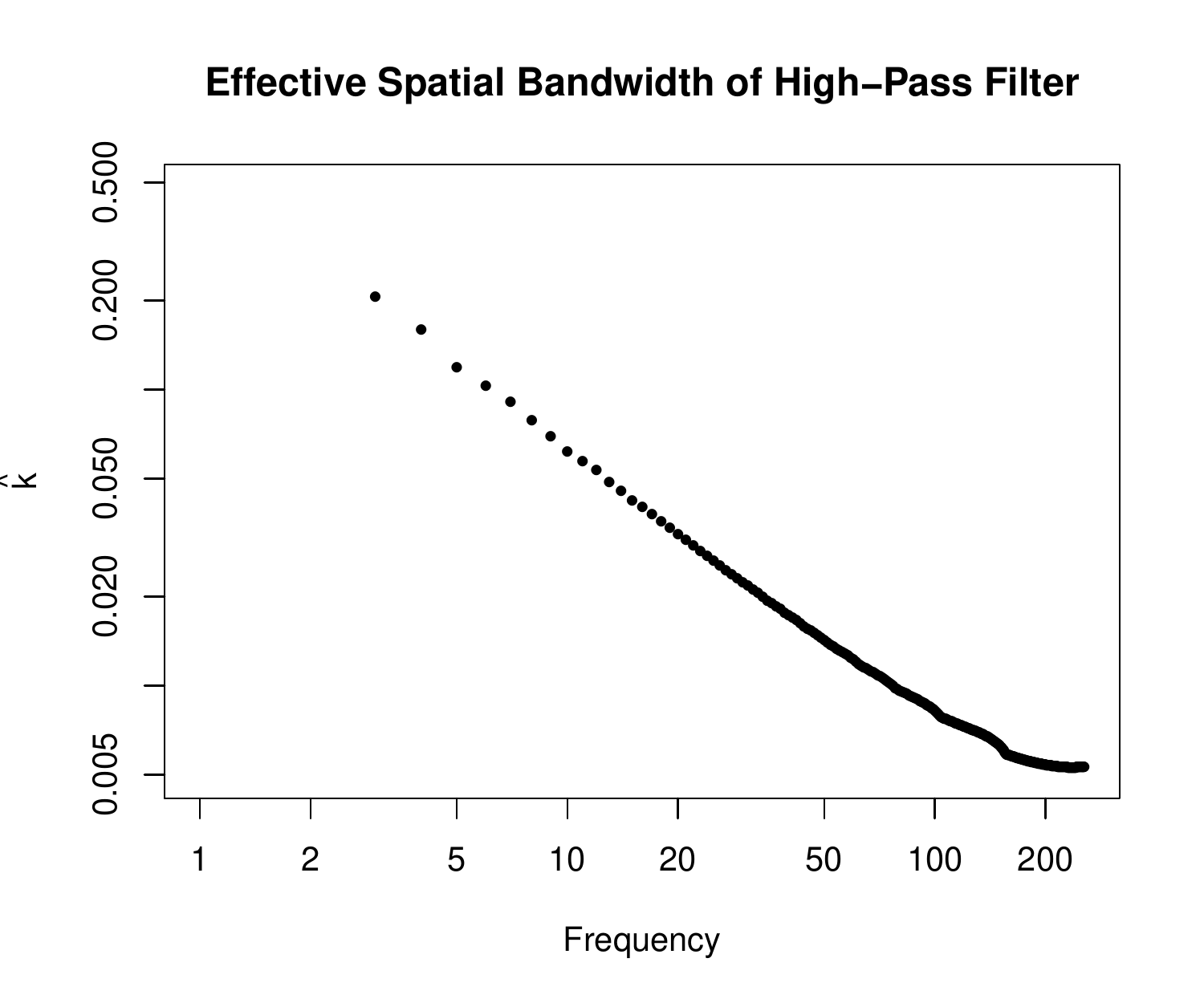}
}
\caption{Effective bandwidth (a) by df for TPRS and (b) by frequency for HPF on a $512 \times 512$ grid over the unit square.}
\label{fig:eff_band_unit_square}
\end{center}
\end{figure}

 The effective bandwidths $\hat k$ for TPRS and a high-pass filter on the unit square are shown in Figure~\ref{fig:eff_band_unit_square}. The difference in ranges of $\hat k$ between the two bases is clear, with the high-pass frequency filter spanning a much larger range of spatial scales than TPRS. The slight ``hiccup'' in Figure~\ref{fig:eff_bandwidth_tprs_unit_square} at $df= 10$ is an artifact of the shape of that particular basis function (and the constraints of the gridded locations). For the high-pass filter, the spatial smoother corresponding to high-pass filters $F_1$ and $F_2$ never cross zero, which is why points are only shown for $\omega \ge 3$. For both adjustment bases, there is an approximately linear relationship between tuning parameter ($df$ or $\omega$) and $\hat k$ on the log-log scale.

 \subsection{Simulation Results}
\begin{figure}[tb]
\begin{center}
\includegraphics[width=\textwidth]{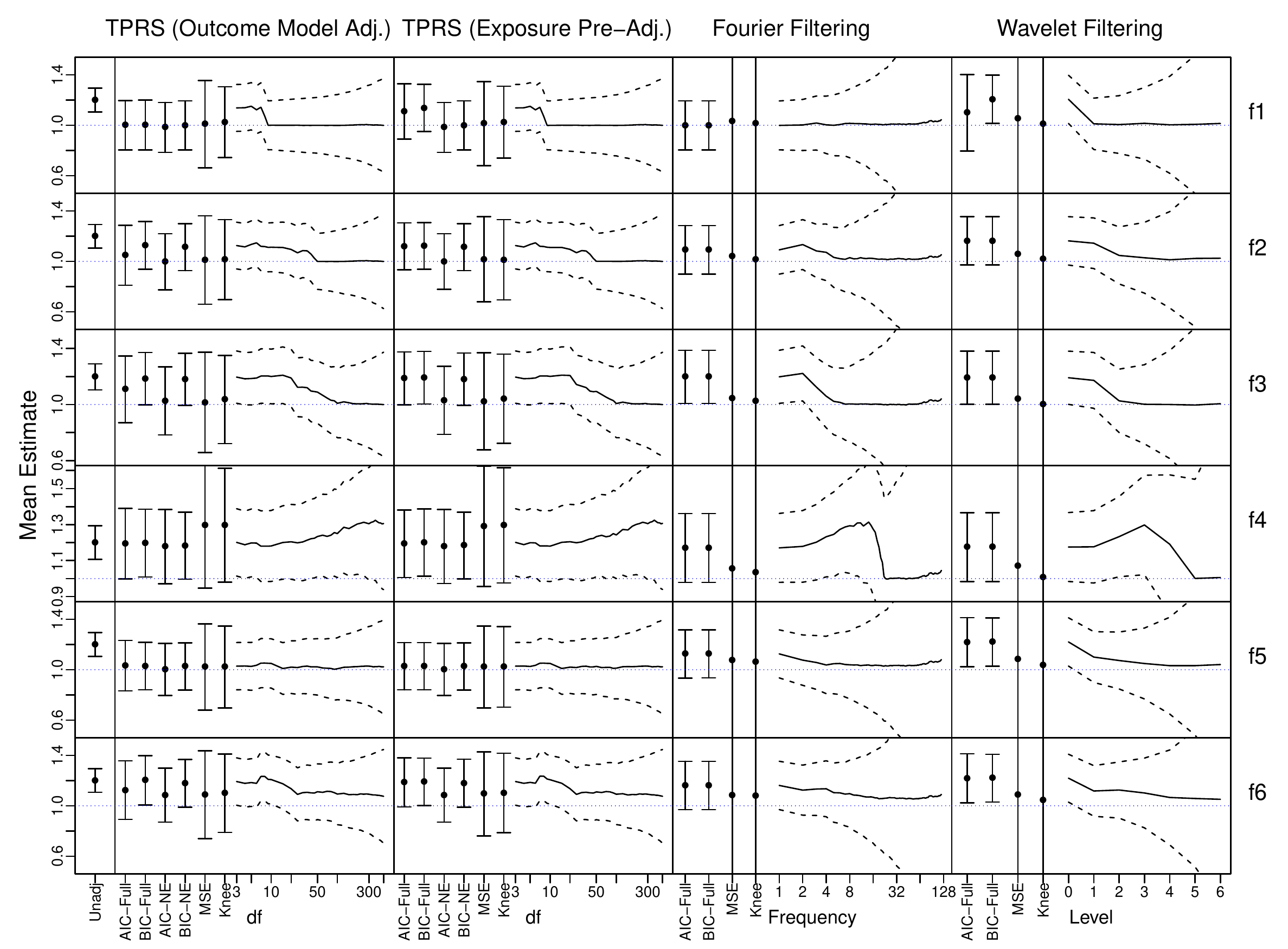}
\caption{Estimates (\drawline{thick})  of $\beta$ in Simulation 1 when pre-adjusting exposure with different choices of spatial basis (panel columns) and  different underlying confounding surfaces (panel rows). The far left column shows the unadjusted estimate. The dashed lines (\drawline{dashed}) and error bars indicate 2x the standard error. The true parameter value $\beta =1$ is plotted as a dotted line (\drawline{thick, blue,densely dotted}).}
\label{fig:sim_S1_g1_results}
\end{center}
\end{figure}

\begin{figure}[tb]
\begin{center}
\includegraphics[width=\textwidth]{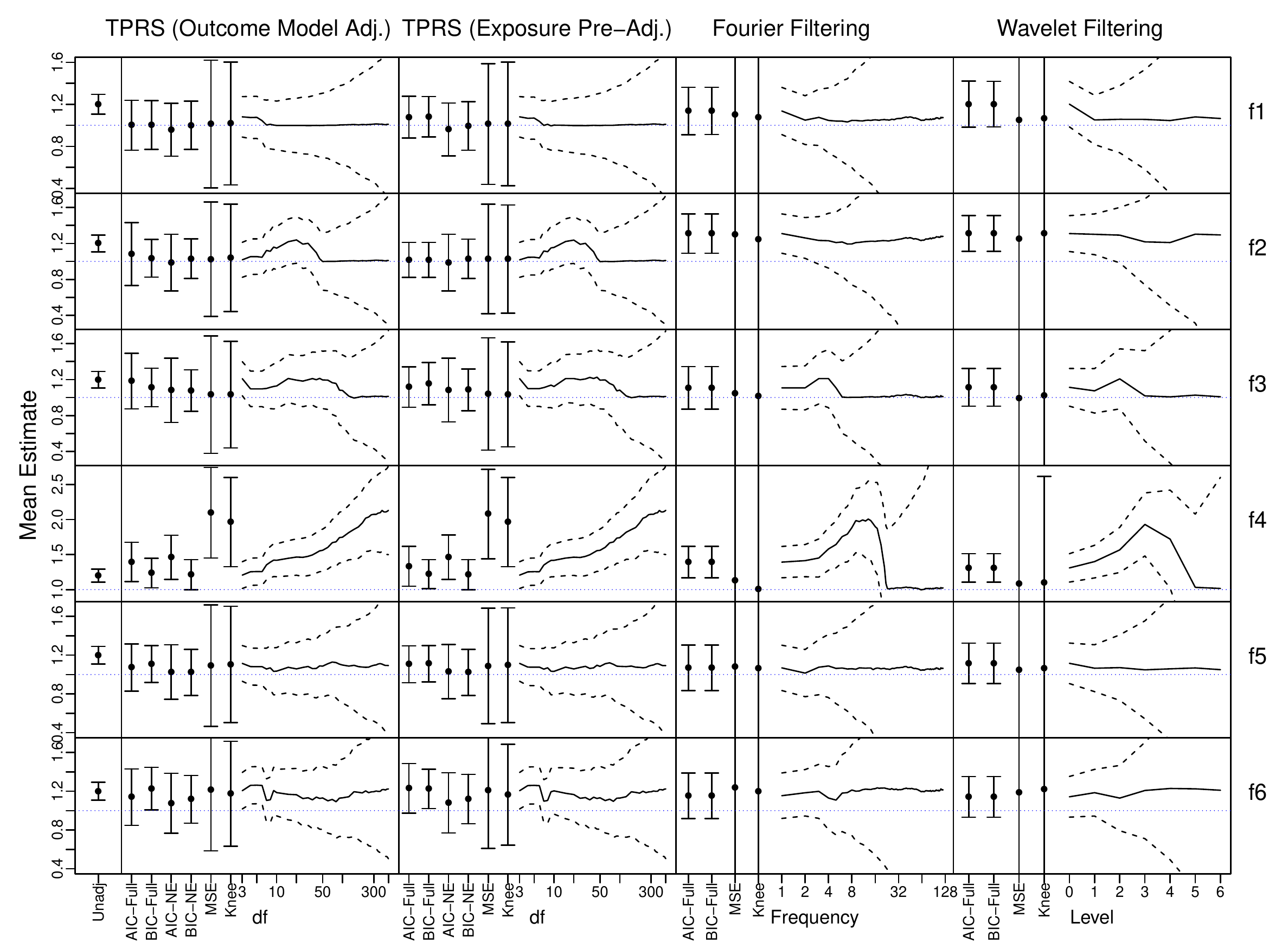}
\caption{Estimates (\drawline{thick})  of $\beta$ in Simulation 2 when pre-adjusting exposure with different choices of spatial basis (panel columns) and  different underlying confounding surfaces (panel rows). The far left column shows the unadjusted estimate. The dashed lines (\drawline{dashed}) and error bars indicate 2x the standard error. The true parameter value $\beta =1$ is plotted as a dotted line (\drawline{thick, blue,densely dotted}).}
\label{fig:sim_S1_g2_results}
\end{center}
\end{figure}

 The point estimates for the different choices of adjustment basis and unobserved confounder surfaces are provided in 
Figure~\ref{fig:sim_S1_g1_results} for Simulation 1 (exposure with fine scale variation) and Figure~\ref{fig:sim_S1_g2_results} for Simulation 2 (exposure with larger scale variation).  In each figure, each panel shows the mean estimate for the estimators that automatically select $\hat m$ and across a range of fixed values of $m$. For both simulations, the bias of the unadjusted estimator that ignores the unmeasured confounding is 1.2, by construction.

 \subsubsection{Results for fixed $m$}
 
When the adjustment approach matches the basis used to generate the confounder, we see that the bias can be removed after sufficient adjustment. For Simulation 1 when confounding is due to $f_1$ and $f_2$, this occurs when there is TPRS adjustment of at least  $df=10$ and $df=50$, respectively. In addition to leading to low bias (top-left panels of Figures~\ref{fig:sim_S1_g1_results}), this leads to low MSE that matches or beats all other estimators (Supplemental Materials Table~1), and nearly correct coverage rates (94\% for nominal 95\% confidence interval [CI]; see Supplemental Materials Table~2).
For $f_3$ and $f_4$, complete adjustment occurs when there is a high pass filter with cutoff at a frequency of 7 and 23, respectively. However the variance of the estimates from the Fourier filtering approach increases greatly with the amount of adjustment, so the MSE remains relatively large despite the lack of bias (Supplemental Materials Table~1).

When the adjustment approach does not match the basis generating the confounder, we see that in some cases most bias can still be eliminated but in other cases bias can persist. 
When the unmeasured confounder has large scale variation ($f_1$, $f_3$, and $f_5$), we see that all approaches are able to eliminate most bias after sufficient adjustment.
However, for the confounder surfaces that have small scale variation, we see the persistence of bias regardless of the amount of adjustment ($f_6$) and bias amplification with increasing adjustment ($f_4$). This bias amplification occurs even when the correct adjustment basis is chosen, which can seen by the increase in bias for confounder surface $f_4$ when adjusting with Fourier filtering up to $\omega = 22$.

In general, we see that TPRS can successfully adjust for confounding even when the unobserved confounder is generated from a different basis. For example, when $f=f_3$,  adjustment with TPRS is able to remove the confounding bias for $df>100$. However, when $f=f_4$, the bias is not eliminated even for very high choices of $df$ relative to the sample size of $n=2,000$. This difference is due to the smaller range of  effective bandwidths of TPRS compared to those of the high-pass filter approach (Figure~\ref{fig:eff_band_unit_square}). By construction, variation in $f_2$ and $f_3$ extends up to $\omega =\sqrt{40} \approx 6.3$ and $\omega = \sqrt{500} \approx 22.4$, respectively. Using the relationship in Figure~\ref{fig:eff_bandwidth_hpf_unit_square}, this corresponds to values of $\hat{k}$ of approximately 0.12 and 0.03, respectively. A value of $df = 85$ corresponds to approximately $\hat k = 0.12$, but there is not value of $df$ for which TPRS can achieve $\hat k = 0.03$.

For Simulation 2, in which the exposure has larger scale spatial variation, we see similar trends to most of the Simulation 1 results. However, we do see greater effects of bias amplification, which now appears when the confounder is $f_2$ as well as when the confounder is $f_4$ (Figure~\ref{fig:sim_S1_g2_results}). Unlike in Simulation 1, the bias induced by  $f_6$ is not reduced at all.

In both sets of simulations, empirical Type I error when $\beta = 0$ followed the same patterns as the coverage rates when $\beta = 1$. Specifically, the error rates were at or close to 0.05 when the correct amount of adjustment was done using the correct basis (Supplemental Materials Table~3).

 \subsubsection{Results when selecting $m$ automatically}

The estimators that automatically select $\hat m$ using information criteria perform generally well.  For adjustment using TPRS in the outcome model in Simulation 1, the model selected by minimizing AIC from the outcome model without exposure (AIC-NE) performs the best in almost all settings. For confounders other than $f_4$, a large amount of adjustment is selected, which leads to reduced bias (Figure~\ref{fig:sim_S1_g1_results}). The MSE of the estimator with adjustment selected by AIC-NE is smaller than the MSE for estimators using  fixed values of $m$ (Supplemental Materials Table~1). For some surfaces, the amount of adjustment selected by BIC-NE yielded slightly smaller MSE, but coverage rates were always better for selection by AIC-NE (Supplemental Materials Table~2). 
In Simulation 2, selection using BIC-NE  performs best in bias (Figure~\ref{fig:sim_S1_g2_results}) and  MSE (Supplemental Materials Table~4)  for all settings except when the confounder is $f_2$. This reflects the benefit of greater penalization for model complexity (which leads to less adjustment) in settings with extensive bias amplification. When the confounder is $f_2$, selection using BIC from the full outcome model performs the best.  However, coverage rates are slightly better using AIC-NE for selection in most cases.

When pre-adjusting with TPRS at observed locations in Simulation 1, the amount of adjustment selected by minimizing AIC-NE was also better in most settings. Again while MSE was slightly lower for the estimator with amount of adjustment selected by BIC-NE, coverage rates were always better for selection by AIC-NE. Performance in all settings was similar to using adjustment for TPRS in the outcome model.
  In Simulation 2, minimizing BIC-NE does better in most settings, having lower variance since it tends to select a smaller model. 

For adjustment using Fourier or wavelet filtering, the minimally-adjusted model is almost always selected when minimizing AIC and BIC since the number of parameters estimated for filtering increases exponentially. This results in lower MSE and better coverage than the unadjusted models, but poorer performance than adjusting with TPRS (Supplemental Materials Tables~1 through 5). The exception is in Simulation 2 for confounders $f_2$ and $f_4$, when the models selected by AIC and BIC do much worse than the unadjusted model.

Selecting the amount of adjustment by the estimated-MSE criterion (equation \eqref{eq:pick_m_mse}) reduced most of the bias in Simulation 1. However, it let to standard errors that were much larger than the other approaches, as can be seen in Figure~\ref{fig:sim_S1_g1_results}. For Fourier and wavelet filtering, the standard errors were so large that a single estimate provided little information. The corresponding MSE was much larger than all other estimators (Supplemental Materials Table~1). In Simulation 2, the bias amplification led to estimates that were as bad or worse.
The \emph{ad hoc} ``knee'' based approach (equation \eqref{eq:pick_m_knee}) performed similarly poorly to the estimated-MSE approach, although had better coverage.

\subsection{Additional Simulations}
We conducted additional simulations that included measured confounders $\bm z$, and the results were not substantively different from Simulation 1 and 2 so are not shown here.
We also conducted simulations in which the number of distinct locations was smaller than the number of subjects, leading to repeated locations. However, little difference was observed between those results compared to the primary results presented here. Approaches to reweighting had some benefit in terms of reducing variance in settings with extreme bias (small variance in $\epsilon$ and $\theta$ chosen for large relative bias), but no practical impact on the amount of bias.

\subsection{Simulation Conclusions}
\label{sec:sim_conclusions}
Based on the results of Simulation 1 and 2, the best overall approach to reducing confounding bias is adjustment with TPRS at observed locations, either in the outcome model or using exposure pre-adjustment. The best approach for selecting the amount of adjustment is to  minimize AIC-NE or BIC-NE (AIC or BIC from an outcome model without the exposure). In cases where bias amplification may be a concern, due to either only large scale variation in the exposure surface or very fine scale confounding, preference should be given to BIC-NE over AIC-NE. 


\section{Sister Study Application}
\label{sec:sisterfullapp}
\begin{figure}[t]
\begin{center}

\subfloat[\label{fig:sister_obsgrid}]{
\includegraphics[width=0.47\textwidth]{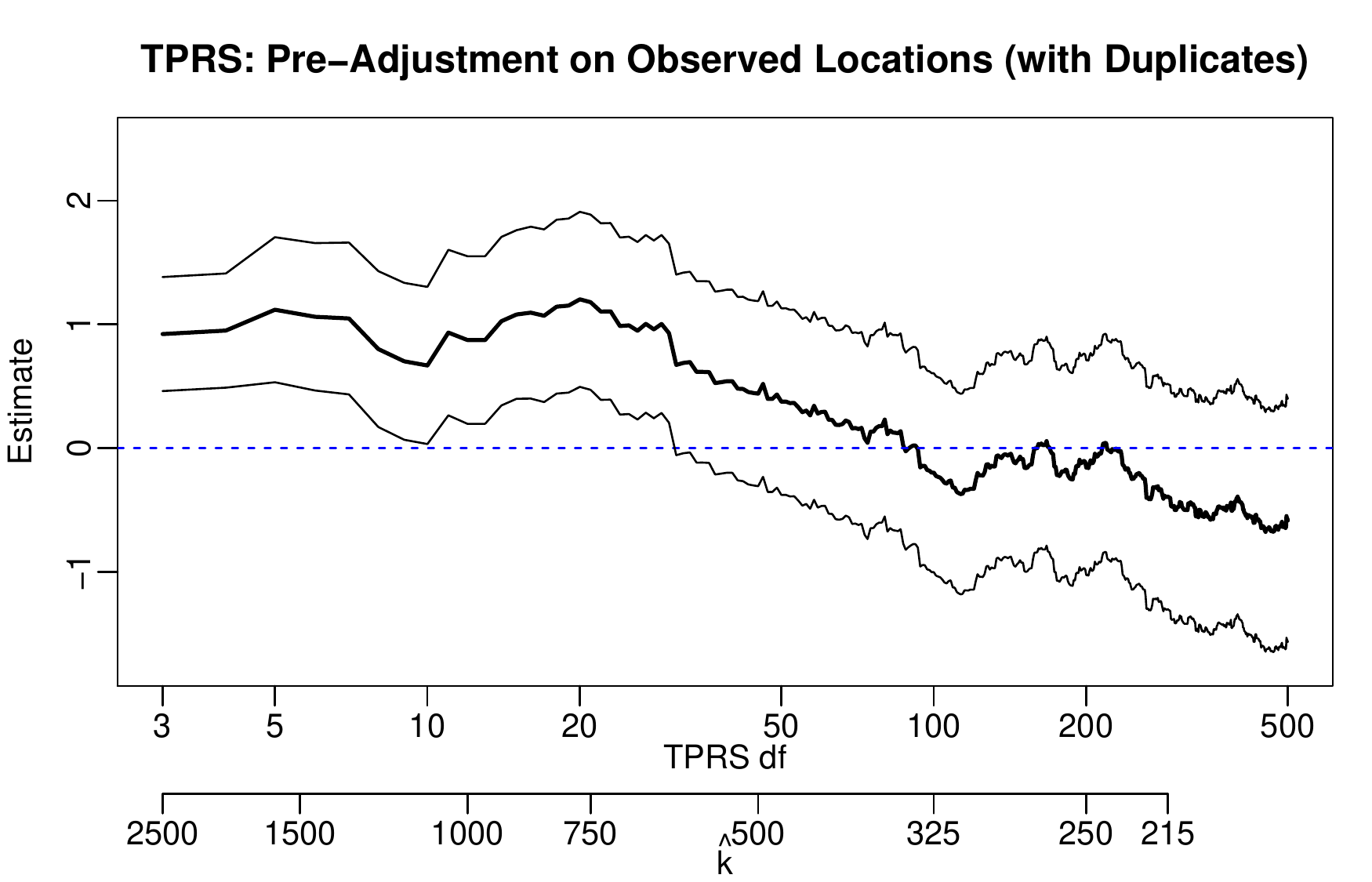}
}
\subfloat[\label{fig:sister_obs1grid}]{
\includegraphics[width=0.47\textwidth]{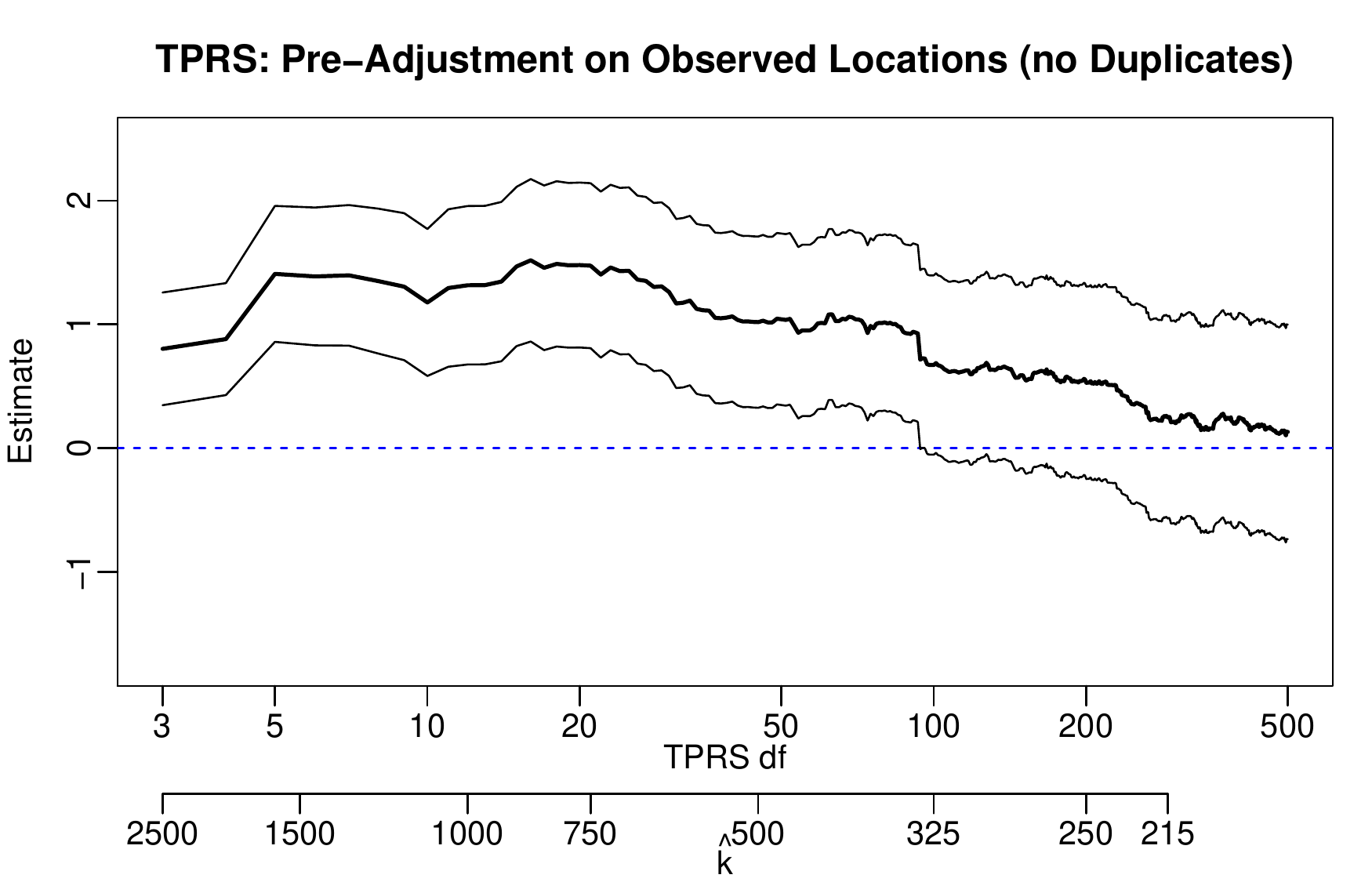}
}\\
\subfloat[\label{fig:sister_tprs_grid}]{
\includegraphics[width=0.47\textwidth]{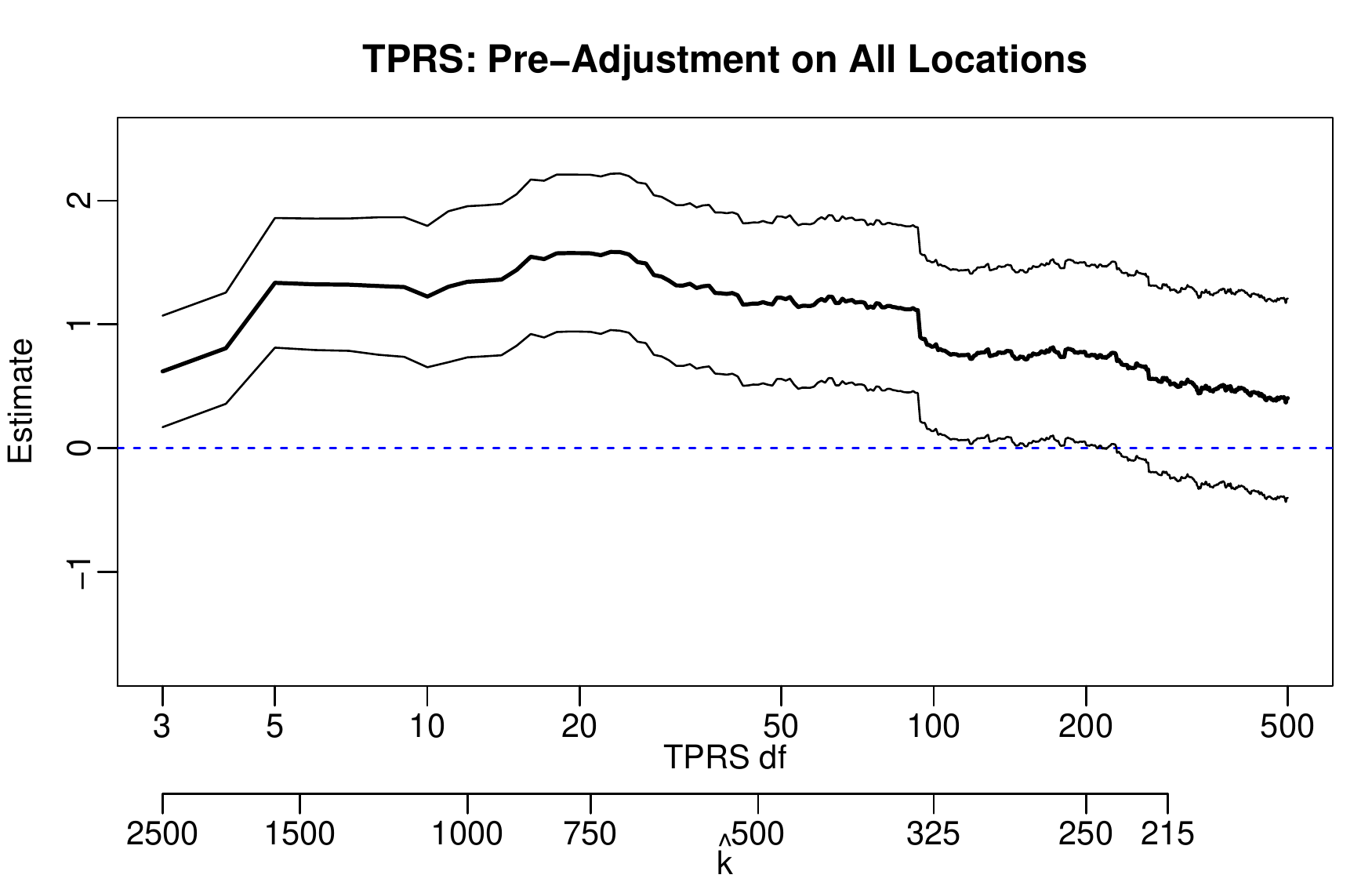}
}
\caption{Estimates (\drawline{ultra thick})  and pointwise confidence intervals (\drawline{black}) of the association between SBP and PM$_{2.5}$ for different amounts of confounding adjustment using TPRS.  The horizontal line (\drawline{blue, dashed}) is at zero.}\label{fig:sister_tprs}
\end{center}
\end{figure}

We now compare these different spatial confounding adjustment approaches in the Sister Study example of the association between SBP and PM$_{2.5}$.
The estimates from performing exposure pre-adjustment using TPRS across all observed grid locations, with each location repeated for the number of subjects at the same location, are provided  
 in Figure~\ref{fig:sister_obsgrid}.
These results are similar to those in Figure~\ref{fig:sister_tprsdf}, as the adjustment basis is the same and the pre-adjustment that leads to Figure~\ref{fig:sister_obsgrid} repeated locations as needed.  Based upon the simulations (Section~\ref{sec:sim_conclusions}), the preferred method for selecting the appropriate amount of adjustment is to minimize AIC-NE  or BIC-NE. Because the exposure surface in this application is quite coarse (25km grid), we use BIC-NE for selecting our primary result to reduce the impact of bias amplification from over-adjustment. This choice results in adjustment using $m=3$ TPRS basis functions. When adjustment is done in the health model, the estimated association is a difference of 0.96 mmHg (95\% CI: 0.50, 1.42) in SBP for each 10 $\mu g/m^3$ difference in PM$_{2.5}$ (Table~\ref{tab:sister_tprs_res}). When the exposure is pre-adjusted, the estimated association is a difference of 0.92 mmHg (95\% CI: 0.46, 1.41) in SBP for each 10 $\mu g/m^3$ difference in PM$_{2.5}$.

 \begin{table}
\caption{\label{tab:sister_tprs_res}Point estimates and 95\% confidence intervals of the difference in SBP (in mmHg) associated with a difference of 10 $\upmu$g/m$^3$ in PM$_{2.5}$ exposure in the Sister Study, corresponding  to different scales of adjustment using TPRS.}
\centering
\fbox{
\begin{tabular}{l c c c c cc}
&  \multicolumn{2}{c}{Outcome Model}  & \multicolumn{2}{c}{Pre-Adjustment}   \\
&  \multicolumn{2}{c}{Adjustment} &\multicolumn{2}{c}{Observed Locations} \\
& $m$ & $\hat \beta$ (95\% CI) &  $m$ & $\hat \beta$ (95\% CI) \\
  \hline
    BIC-NE & 3 & $0.96$ ($0.50$, 1.42) & 3&0.92 (0.46, 1.41) \\
  AIC-NE & 401 & $-0.49$ ($-1.44$, 0.45) & 401  & $-0.43$ ($-1.38$, 0.52) \\
        BIC & 3 & $0.96$ ($0.50$, 1.42) & 3 &0.92 (0.46, 1.41) \\
  AIC & 401 & $-0.49$ ($-1.44$, 0.45) & 4  &0.95 (0.49, 1.41) \\
  $\hat k = $ 1,000 km & 12 &  1.05 (0.37, 1.73) & 12 &  0.87 (0.20, 1.55) \\
MSE & 352 & $-0.59$ ($-1.52$, 0.33) & 113 & $-0.37$ ($-1.18$, 0.44) \\
Knee & 11&  $1.10$  (0.43, 1.78) &  11 & $0.93$ ($0.26$, 1.60)  \\
\hline
\hline
& \multicolumn{2}{c}{Pre-Adjustment}  & \multicolumn{2}{c}{Pre-Adjustment} \\
& \multicolumn{2}{c}{Observed Locations (1 time  each) } &  \multicolumn{2}{c}{All Locations (1 time each)}\\
& $m$ & $\hat \beta$ (95\% CI) &  $m$ & $\hat \beta$ (95\% CI) \\
  \hline
AIC & 4 & 0.88 (0.43, 1.33) & 5 & 1.34 (0.81, 1.86)\\
BIC & 3 & 0.80 (0.35, 1.26) & 4 & 0.81 (0.36, 1.26) \\
$\hat k = $1,000 km &   12 & 1.32 (0.68, 1.96) & 12 &  1.34 (0.73, 1.95) \\
MSE & 268 &0.23 ($-0.58$, 1.03) & 3 & 0.62 (0.17, 1.07) \\
Knee &  5 & 1.41 (0.86, 1.96) & 5 & 1.34 (0.81, 1.86) \\
\end{tabular}}
\end{table}

For comparison, Table~\ref{tab:sister_tprs_res} also includes the results for the other approaches to selecting $m$. Using AIC-NE or AIC  results in a large amount of adjustment being selected ($m=401$). This appears to be extensive overfitting of the model; the increases in the value of the log-likelihood as $m$ increases are small relative to the large sample size ($n=47,206$). The MSE approach also chooses a large amount of adjustment: $m=352$ for outcome model adjustment and $m=113$ for exposure pre-adjustment. All of these approaches with large adjustment result in negative point estimates due to the downward trend observed in Figures~\ref{fig:sister_tprsdf} and \ref{fig:sister_obsgrid}. This downward trend is likely due to bias amplification from residual confounding, since PM$_{2.5}$ only explains a small fraction of the variation in SBP and the measured confounders, while numerous, likely cannot capture all of the confounding relationships. Furthermore, the coarse spatial scale of the exposure means that variation due to unmeasured confounders is likely finer-scale  than the exposure.

Results from two alternative approaches to exposure pre-adjustment are presented in Figures~\ref{fig:sister_obs1grid} and \ref{fig:sister_tprs_grid}. These figures correspond, respectively, to pre-adjustment at observed locations, with duplicated locations removed, and pre-adjustment over all locations in the domain, with each location included once. 
The point estimates corresponding to the different selection methods for these adjustment approaches are provided in the bottom half of Table~\ref{tab:sister_tprs_res}. The amount of adjustment selected by AIC and BIC (using the full outcome model) is small, ranging from $m=3$ to $m=5$. These choices of adjustment yield point estimates ($0.80$ to $1.34$) similar to the main results from the models that adjust at all observed locations ($0.96$ and $0.92$).
The difference in estimates between Figures~\ref{fig:sister_obs1grid} and \ref{fig:sister_tprs_grid} is relatively small, suggesting that the restriction of the pre-adjustment to the observed locations does not have a large impact on the results. However, there is a notable qualitative difference between these results and those that included duplicate locations for the adjustment (Figures~\ref{fig:sister_tprsdf} and \ref{fig:sister_obsgrid}). While all four approaches show a negative trend in the point estimates for large values of $m$, the approaches that adjust using observed locations including duplicates decrease at smaller values of $m$ and yield negative point estimates.  
We explored using reweighting to correct for the impact of the duplicated locations in the pre-adjustment approaches and obtained results that were either qualitatively similar to the results without reweighting or were highly unstable (see Supplemental Materials Section C.2). The magnitude of the difference in the point estimates from these alternative approaches using TPRS (including or excluding duplicate locations) suggests that residual confounding is likely occurring in this setting.

Estimates for a fixed choice of $\hat k = $ 1,000 km are also provided in  Table~\ref{tab:sister_tprs_res}. This corresponds to adjustment using $12$ degrees of freedom (Supplemental Materials Section C.1).
 A bandwidth of this size smooths  large scale variation across the contiguous United States (which is approximately 4,500 km by 2,900 km). Because there are well-established large-scale trends in health outcomes across the United States \citep{Mensah2005}, this value was chosen so that the analysis accounts for long-scale trends in systolic blood pressure.

\begin{figure}[t]
\begin{center}
\subfloat[\label{fig:sister_hpf}]{
\includegraphics[width=0.47\textwidth]{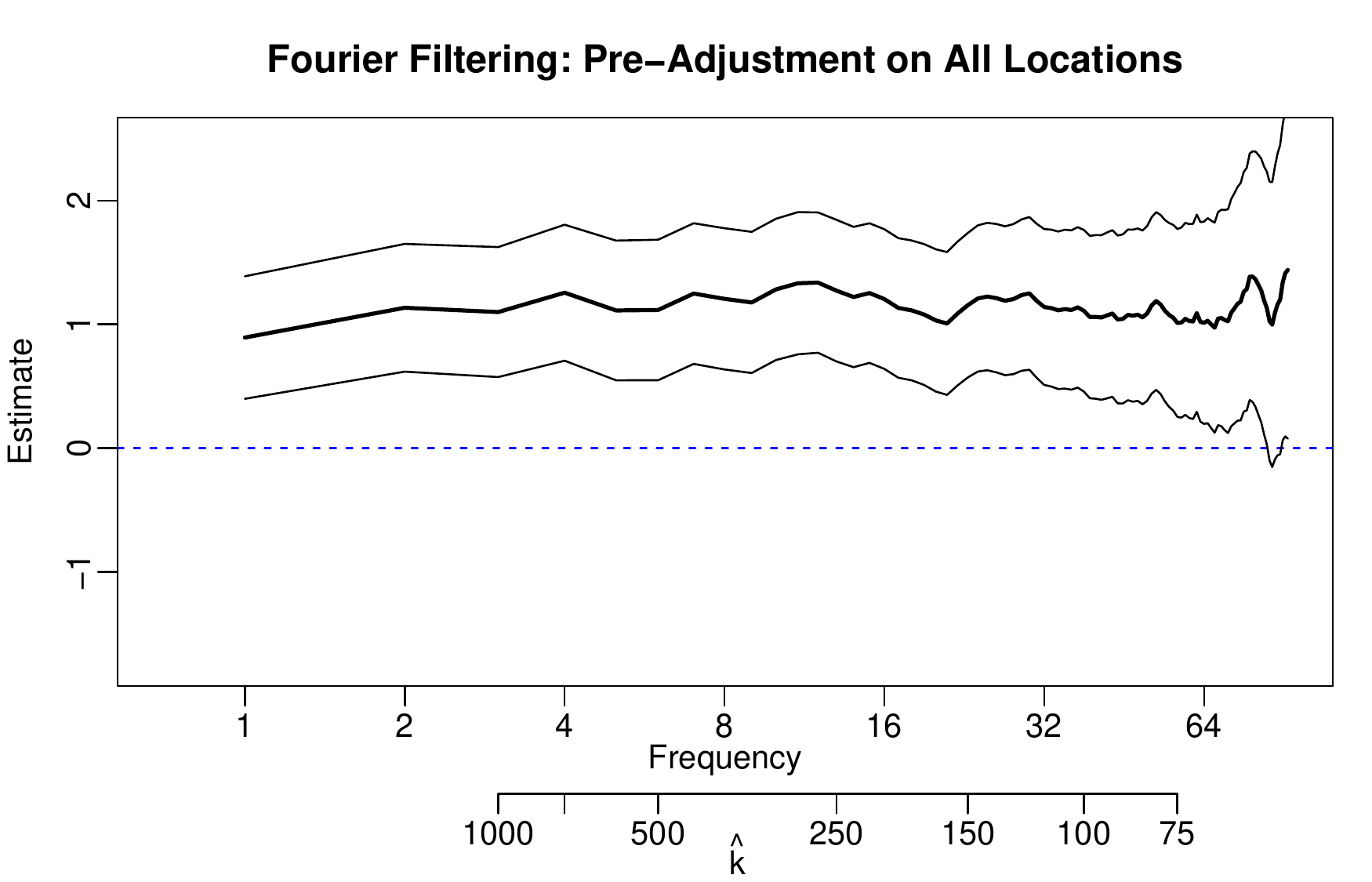}
}
\subfloat[\label{fig:sister_wave}]{
\includegraphics[width=0.47\textwidth]{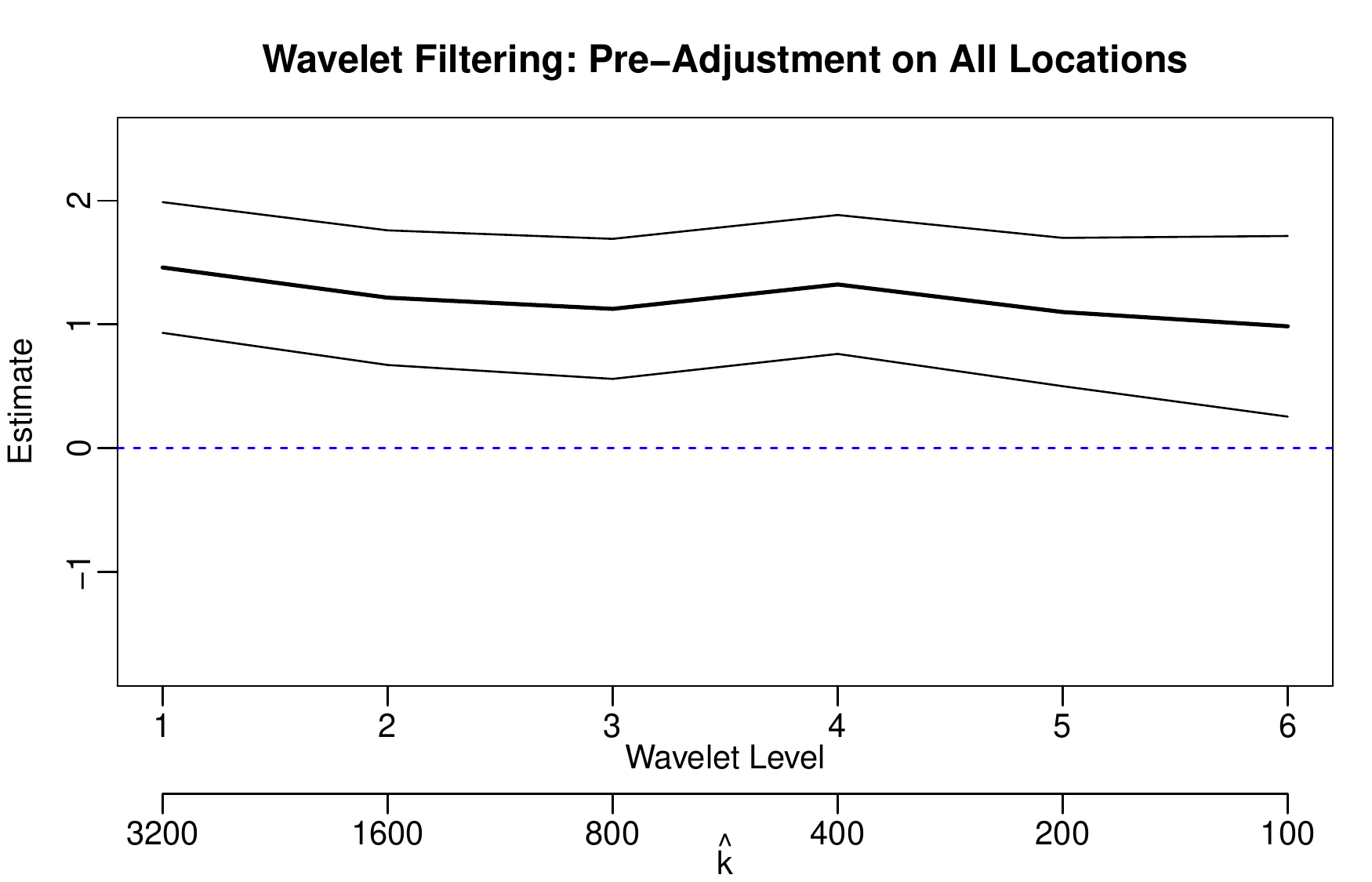}
}
\caption{Estimates (\drawline{ultra thick})  and pointwise confidence intervals (\drawline{black}) of the association between SBP and PM$_{2.5}$ for different amounts of confounding adjustment using Fourier and wavelet filtering. The horizontal line (\drawline{blue, dashed}) is at zero.}%
\label{fig:sister_all}
\end{center}
\end{figure}

Based on the simulation results and duplicated observed locations, we recommend adjustment using TPRS as described above instead of using the Fourier or wavelet approaches. However, we present results for those approaches here for illustration.
To apply the pre-adjustment approaches with Fourier and wavelet basis functions, we embed the gridded locations within a larger square grid.  
Because the predictions of 2006 annual average  PM$_{2.5}$  of \citet{Sampson2013} are only defined over the contiguous United States, 
the added points are assigned exposure concentration of zero. This results in a grid of size 184 $\times$ 184 for the Fourier approach and 256 $\times $ 256 for the wavelet approach.  We see from Figures~\ref{fig:sister_hpf} and \ref{fig:sister_wave} that the point estimates are relatively stable for all amounts of adjustment. This is probably due in part to the sinusoidal and wavelet basis functions not being representative of the spatial structure of the real-world unmeasured confounders, so increased adjustment removes little variation from the outcome or exposure. Additionally, because these adjustment methods rely on each location being present only a single time, we expect trends similar to Figures~\ref{fig:sister_obs1grid} and \ref{fig:sister_tprs_grid}.

\section{Discussion}
\label{sec:discussion}
We have examined different approaches to adjusting for unmeasured spatial confounding and quantifying and selecting the scale of adjustment.  These approaches are motivated as extensions of  time series methods for temporal confounding \citep{Peng2006,SzpiroSheppard2014}. 
We presented a method for comparing the spatial scales across choices of basis using the effective bandwidth $\hat k$. We showed examples using  TPRS, a Fourier basis, and wavelets, each of which indexes variability in different forms and on different scales.
We showed that adjustment with TPRS is limited to a smaller range of spatial scales than the Fourier approach. However, the smaller bandwidth range of TPRS can have more flexibility at these scales. But when the true variation is beyond the scales TPRS can reach, it fails to remove confounding bias.

For a particular application, a choice of the amount of adjustment must be made. We identified selecting the amount of adjustment using AIC-NE and BIC-NE as the preferred approaches for most settings.
When there is substantive knowledge available about the scales of variation involved, $\hat k$ can be chosen \emph{a priori}.
\emph{Ad hoc} approaches such as the ``knee'' and estimated MSE methods performed poorly.

Pre-adjusting the exposure by Fourier filtering or wavelet thresholding is attractive due to the orthogonality of those bases. However, they are both severely limited by the requirement of a square grid, assumptions of periodicity, and the need to reweight the population. For wavelets, the thresholding at dyadic levels also leads to limited available intervals for adjustment. Although in settings where there is a priori knowledge about confounding relationships,  non-uniform thresholding of the wavelet coefficients could add flexibility. 
For these reasons, the more flexible TPRS are likely to be preferred over the Fourier or wavelet basis for settings where TPRS can represent variation at the scale of interest, such as the example study of PM$_{2.5}$ in the Sister Study.
The Fourier basis ccould be chosen when very-fine scale adjustment is needed or there is external evidence to support periodic variation in the confounders.

One of the key results from our simulations is that the addition of spatial basis functions, or an equivalent pre-adjustment procedure, does not necessarily reduce bias in the point estimate.  In fact, it can lead to substantial bias amplification. 
This is an important contrast to the time series settings, in which more aggressive adjustment is recommended to reduce bias from unmeasured confounding \citep{Peng2006}. The exposure in a time series settings are typically measured directly, which means that there is substantial residual variation in the exposure at each time point, whereas the use of an exposure prediction model in a spatial context means that spatial exposures are often quite smooth.
The potential for bias amplification means that using changes in the point estimates is often not a good approach to assessing the extent of confounding bias.

We have presented results here in the context of a linear health model. However, these approaches to spatial confounding adjustment can be applied in generalized linear model settings. The patterns of bias reduction (or increase) for a sequence of adjustment values $m$ will differ by context and choice of link function, but the connection between adjustment basis, spatial scale, and overall interpretation remains the same as the linear case. For example, \cite{Keet2018} used TPRS to adjust for large-scale confounding across the contiguous United States in an analysis of particulate matter and asthma-related outcomes. Automated selection could be done using extensions of information criteria such as QIC \citep{Pan2001}.

In summary, we have presented  methods for describing and selecting the spatial scale of spatial confounding adjustment using different choices of basis. 
These methods can be used to more accurately describe the extent of spatial confounding adjustment in future studies with spatial exposures.

\section*{Acknowledgements}
This work was supported by grants T32ES015459 and R21ES024894 from the the National Institute for Environmental Health Sciences (NIEHS). The Sister Study supported in part by the Intramural Research Program of the NIH, NIEHS (Z01ES044005). This work was also supported in part by the US Environmental Protection Agency (EPA) through awards RD83479601 and RD835871. This was work has not been formally reviewed by the EPA. The views expressed in this document are solely those of the authors and do not necessarily reflect those of the Agency. EPA does not endorse any products or commercial services mentioned in this publication.

\bibliographystyle{chicago}
\bibliography{Keller_Spatial_Confounding}

\end{document}